\documentclass[useAMS,usenatbib]{mn2e}

\usepackage{graphicx}
\usepackage{natbib}
\newcommand{\HI}{H{\,\small I}}
\newcommand{\ltsima} {$\; \buildrel < \over \sim \;$}
\newcommand{\gtsima} {$\; \buildrel > \over \sim \;$}
\newcommand{\lta} {\lower.5ex\hbox{\ltsima}}
\newcommand{\gta} {\lower.5ex\hbox{\gtsima}}
\newcommand{\kmsMp}{km s$^{-1}$ Mpc$^{-1}$}
\newcommand{\kms}{km~s$^{-1}$}
\newcommand{\FRI}{FR{-\small I}}
\newcommand{\FRII}{FR{-\small II}}
\newcommand{\OIII}{O{\,\small III}}

\title[Large-scale H{\small \ }{\large I} disc around NGC~612]{Enormous disc of cool gas surrounding the nearby powerful radio galaxy NGC~612 (PKS~0131-36)}
\author[B. H. C. Emonts et al.]{B. H. C. Emonts$^{1}$\thanks{E-mail:emonts@astro.columbia.edu}, R. Morganti$^{2,3}$, T. A. Oosterloo$^{2,3}$, J. Holt$^{4,5}$,
\newauthor C. N. Tadhunter$^{4}$, J. M. van der Hulst$^{3}$, R. Ojha$^{6}$, E. M. Sadler$^{7}$\\
$^{1}$Department of Astronomy, Columbia University,  Mail Code 5246, 550 West 120th Street, New York, N.Y. 10027, USA\\ 
$^{2}$Netherlands Foundation for Research in Astronomy, Postbus 2, 7990 AA Dwingeloo, the Netherlands\\
$^{3}$Kapteyn Astronomical Institute, University of Groningen, P.O. Box 800, 9700 AV Groningen, the Netherlands\\
$^{4}$Department of Physics and Astronomy, University of Sheffield, Sheffield S3 7RH, UK\\
$^{5}$Leiden Observatory, Leiden University, P.O. Box 9513, 2300 RA Leiden, the Netherlands\\
$^{6}$United States Naval Observatory /NVI, 2212 40th Pl NW $\#$2, Washington DC 20007, USA\\
$^{7}$School of Physics, University of Sydney, NSW 2006, Australia\\
}
\begin{document}

\date{}

\pagerange{\pageref{firstpage}--\pageref{lastpage}} \pubyear{2007}

\maketitle

\label{firstpage}

\begin{abstract}
We present the detection of an enormous disc of cool neutral hydrogen (H{\,\normalsize I}) gas surrounding the S0 galaxy NGC~612, which hosts one of the nearest powerful radio sources (PKS~0131-36). Using the Australia Telescope Compact Array, we detect $M_{\rm HI} = 1.8 \times 10^{9} M_{\odot}$ of H{\,\normalsize I} emission-line gas that is distributed in a 140 kpc wide disc-like structure along the optical disc and dust-lane of NGC~612. The bulk of the gas in the disc appears to be settled in regular rotation with a total velocity range of 850 \kms, although asymmetries in this disc indicate that perturbations are being exerted on part of the gas, possibly by a number of nearby companions. The H{\,\normalsize I} disc in NGC~612 suggests that the total mass enclosed by the system is $M_{\rm enc} \approx 2.9 \times 10^{12}\ {\rm sin}^{-2}i\ M_{\odot}$, implying that this early-type galaxy contains a massive dark matter halo. We also discuss an earlier study by Holt et al. that revealed the presence of a prominent young stellar population at various locations throughout the disc of NGC~612, indicating that this is a rare example of an extended radio source that is hosted by a galaxy with a large-scale star-forming disc. In addition, we map a faint H{\,\normalsize I} bridge along a distance of 400 kpc in between NGC~612 and the gas-rich ($M_{\rm HI} = 8.9 \times 10^{9} M_{\odot}$) barred galaxy NGC~619, indicating that likely an interaction between both systems occurred. From the unusual amounts of \HI\ gas and young stars in this early-type galaxy, in combination with the detection of a faint optical shell and the system's high infra-red luminosity, we argue that either ongoing or past galaxy interactions or a major merger event are a likely mechanism for the triggering of the radio source in NGC~612. This paper is part of an ongoing study to map the large-scale neutral hydrogen properties of nearby radio galaxies and it presents the first example of large-scale H{\,\normalsize I} detected around a powerful Fanaroff $\&$ Riley type{-\normalsize II} (FR{-\normalsize II}) radio galaxy. The H{\,\normalsize I} properties of the FR{-\normalsize II} radio galaxy NGC~612 are very similar to those found for low-power compact radio sources, but different from those of extended Fanaroff $\&$ Riley type{-\normalsize I} (FR{-\normalsize I}) sources. 
\end{abstract}


\section{Introduction}
\label{sec:introduction}

Early-type galaxies are often believed to be gas poor systems. However, recent studies show that the majority of nearby field early-type galaxies contain modest amounts of neutral hydrogen (\HI) gas \citep{mor06b}, while at least $6-15 \%$ of all nearby early-type galaxies even contain \HI\ masses similar to that of the Milky Way\footnote{$M_{\rm \HI,\ \rm MW} = 5 \times 10^9 M_{\odot}$ \citep{hen82}.}, often distributed in large-scale rotating discs and rings \citep{sad02,oos07}. Early-type galaxies in clusters appear to be much more devoid of \HI\ gas, as suggested by a recent \HI\ survey of the VIRGO cluster by \citet{dis07}. Nevertheless, these recent results show that the gaseous component can be an important factor to consider in the hierarchical picture of galaxy formation.

Although N-body and hydrodynamic simulations investigate in detail the stellar component in the formation of early-type galaxies \citep[e.g.][]{del06}, the fate of the gas is much less clear. This is on the one hand related to the complicated feedback processes (such as star formation and outflow phenomena) that are effective on the gas, while on the other hand the processes of gas accretion are not yet well understood. The last factor being subject to a large range of possible accretion mechanisms, from the merging of two equal mass galaxies \citep[e.g.][]{bar02} or a more gradual infall of small gas-rich companions \citep[e.g.][]{hul05} to perhaps slow accretion through the cooling of primordial hot circum-galactic gas \citep[e.g.][]{ker05}.

Radio galaxies form an interesting sub-set in understanding the gaseous properties of early-type galaxies, because the triggering mechanism of radio sources is related to gas-dynamical processes that can remove enough angular momentum from the gas for it to be deposited on to a super-massive black hole in the centre of the galaxy. Various possible triggering mechanisms have been suggested, often for particular types of radio sources. These mechanisms range from cooling flows \citep[e.g.][]{fab94} to galaxy interactions, dry mergers \citep{col95} and major gas-rich collisions and mergers between equal mass galaxies \citep[e.g.][]{hec86,bau92}. The last group is particularly interesting with respect to investigating the hierarchical model of galaxy evolution, because the typical end-product of a major merger  between two gas-rich disc galaxies is an early-type galaxy \citep[e.g.][]{naa99}.

A significant fraction of the more powerful, edge-brightened Fanaroff $\&$ Riley type{-\small II} \citep[\FRII;][]{fan74} radio galaxies show optical peculiarities reminiscent of a gas-rich galaxy collision or merger \citep[such as tidal-tails, -bridges, -plumes, etc.;][]{hec86,smi89,bau92}. Detailed studies of extended emission-line regions around powerful radio galaxies reveal that mergers or interactions are often associated with their early-type host galaxies \citep[e.g.][]{tad89,koe98,ins07,fu07}. In addition, a number of \FRII\ radio galaxies have been studied in great detail to show evidence for the presence of an intermediate age stellar population, likely the result of a merger-induced starburst \citep[e.g.][]{tad05,fu07}. This indicates that powerful radio galaxies are ideal cases to investigate in detail these active phases of galaxy evolution. Furthermore, feedback from Active Galactic Nuclei (AGN) is more and more recognised to be a crucial factor in regulating the gas distribution and, related, the star formation properties of early-type galaxies \citep[e.g][]{dim05,spr05,hop05}. In recent studies, we find that the feedback effects of powerful radio sources on the inter-stellar medium (ISM) result in some cases in significant outflows of mostly neutral hydrogen gas from the central region of the host galaxy \citep[][]{mor03apj593,mor05a,emo05,mor05b}, indicating that the radio-loud phase may have a direct impact on the gaseous properties, and hence the evolution, of the galaxy.

The early-type galaxy NGC~612 is one of the nearest systems that contain a powerful radio source with \FRII\ properties. NGC~612 is therefore is an excellent candidate to investigate in detail the presence and possible distribution of gas and compare this to the origin of the system and triggering of the radio source. In this paper we present results of a study of the neutral hydrogen (\HI) 21cm-line gas in NGC~612 and its environment. The combined spatial and kinematical information obtained with radio synthesis observations of the \HI\ emission-line gas provides an efficient tool to study both the gas distribution and kinematics, as well as to trace, date and classify tidal signatures from galaxy interactions and mergers (tails, bridges, plumes, discs, etc.). 
We compare our \HI\ results with a stellar population analysis of NGC~612 done by \citet{hol07} and with the properties of the radio source. We also compare our findings with interesting results that we recently obtained on the \HI\ content and distribution in lower power (compact and \FRI) radio galaxies \citep{emo07}.\\
\vspace{2mm}\\
{\it NGC~612}\\
\vspace{-2mm}\\
NGC~612 has been studied in detail by \citet{ver01} to be a typical S0 galaxy, which is located at ${\sl z} = 0.0297$.\footnote{For H$_{0} = 71$ \kmsMp\ (used throughout this paper); this puts NGC~612 at a distance of 125 Mpc and 1 arcsec = 0.6 kpc} \citet{gos80} traced a regularly rotating emission-line disc out to a radius of 28 kpc along the major axis of the galaxy\footnote{Corrected for H$_{0} = 71$ \kmsMp}, although the emission-lines are weak \citep{wes66,gos80,tad93}. NGC~612 also has a prominent dust-lane along the disc, almost perpendicular to the radio-axis \citep[e.g.][]{eke78,ver01}. After initial reports by \citet{tad93} about peculiar optical spectral features (strong and narrow absorption features and a weak 4000-\AA\ break) and by \citet{rai05} about the presence of young stellar populations in NGC~612, recently \citet{hol07} traced a young stellar population with age $\sim 0.04 - 0.1$ Gyr throughout the stellar disc of the host galaxy.

The large-scale, two sided structure of the radio source PKS~0131-36 in host galaxy NGC~612 became apparent from radio observations by \citet{eke78} and was studied at higher resolution by \citet{mor93}. The eastern radio lobe has a clear \FRII\ \citep{fan74} morphology, with a bright hot-spot near its outer edge, while the western lobe is somewhat more diffuse. \citet{gop00} classified the radio source as a so called HYMORS (HYbrid MOrphology Radio Source), with the eastern lobe of type \FRII\ and the western lobe of type \FRI. The total radio power is $P_{\rm 4.8 GHz} \sim 0.8 \times 10^{25}$ W~Hz$^{-1}$ \citep[][]{mor93},\footnote{See footnote 3} which is at the border between the typical power of \FRI\ and \FRII\ sources \citep[e.g.][]{owe89,owe91,led02}. No bright optical AGN is apparent in NGC~612, given the weak optical emission-lines \citep{hol07,tad93} and the low percentage of polarised light \citep{bri90} in the nuclear region.


\section{Observations}
\label{sec:observations}

The \HI\ observations of NGC~612 were done with the Australia Telescope Compact Array (ATCA) for 2 $\times$ 12 hrs on 25/26/27 June 2003 using the 750C array-configuration and for 2 $\times$ 12 hrs on 10/11 Oct. 2005 using the EW214 array. Combining these two configurations gives a good {\it uv} coverage, with baselines ranging from 31 to 750m for the inner five antennas of the array. The sixth ATCA antenna, located at a distance of about 5 km from the inner five antennas, was in operation only during the 750C configuration. We used this sixth antenna only for the very high spatial resolution absorption study described below. All observations were done with 16 MHz bandwidth, 512 channels and 2 polarisations.

For the data reduction and visualisation we used the {\small MIRIAD} and {\small KARMA} software. After flagging and calibration, we separated the continuum from the line data in each individual data set by fitting either a first or a second order polynomial to the line-free channels. In order to trace the structure of the radio source in detail, the continuum data obtained with the 750C array (the configuration that provides the higher spatial resolution of the two) were used to create the uniform weighted continuum image shown in Fig. \ref{fig:continuum} [beam = $62.4 \times 38.0$ arcsec$^{2}$; position angle (PA) 0.9$^{\circ}$]. The combined line-data of all the four runs were used to construct a data-cube with robust weighting +1 \citep[see][]{bri95}, beam of 154.25 $\times$ 89.59 arcsec$^{2}$ (PA -1.3$^\circ$), velocity resolution of 27.6 \kms\ (after binning two consecutive channels and subsequently Hanning smoothing the data) and noise level of 0.64 mJy beam$^{-1}$. From this data set a mask was created by smoothing the data spatially by a factor 1.7, applying another Hanning smooth in velocity, and subsequently masking out all the signal below 3$\sigma$. This mask was used to extract a total intensity image of the data by adding the signal above 2.5$\sigma$ in the regions that were not masked out (Fig. \ref{fig:HIemission}).

Finally, a continuum- and a line-data set with the highest possible resolution were constructed from the 750C array data including the sixth ATCA antenna. As described in detail in Section \ref{sec:absorption}, these data sets are not useful for accurate emission-line or continuum studies (because of the gaps in {\it uv} coverage), but they are used for a detailed \HI\ absorption analysis. Using uniform weighting, the beam-size of these high resolution data sets is $7.8 \times 4.6$ arcsec$^{2}$ (PA -1.3$^{\circ}$), with a velocity resolution of 13.8 \kms\ for the line-data.

\subsection{VLT acquisition image}
\label{sec:VLTimage}
 
A short-exposure VLT (Very Large Telescope) acquisition image was taken with the FORS2 (FOcal Reducer and low dispersion Spectrograph 2) instrument at ESO-VLT-U4 on Sept. 25th 2003, using the R-SPECIAL filter. The integration time was 10 seconds and observations were done at an airmass of 1.1. The image consists of two simultaneous exposures at two separate chips, with a small overlap region to mosaic them. We did a bias subtraction and flat-fielding on each exposure before combing them. We then rotated the image to the north-up position shown in Fig. \ref{fig:vlt}.

\section{Results}
\label{sec:results}

\begin{figure}
\centering
\includegraphics[width=0.46\textwidth]{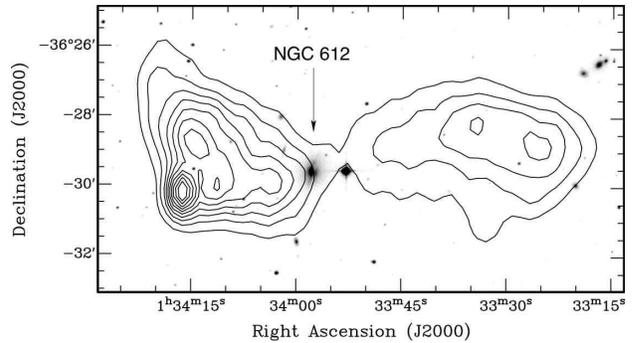}
\caption{Radio continuum map (contours) -- constructed from the 750C array data -- overlaid on to an optical SDSS image of NGC~612 (grey-scale). Contour levels range from 41 to 468 in steps of 47 mJy beam$^{-1}$.}
\label{fig:continuum}
\end{figure}

\subsection{Radio Continuum}
\label{sec:continuum}

Figure \ref{fig:continuum} shows the continuum map of the radio source PKS~0131-36 in NGC~612. The map clearly shows the typical \FRII\ morphology of the eastern lobe (ending in a hot-spot), while the western lobe appears to be more diffuse. The western lobe shows a slight bend, and also the eastern lobe looks somewhat distorted, suggesting that the radio source changed its direction over its lifetime. The total radio power that we derive for this source is $P_{\rm 1.4 GHz} \sim 1.5 \times 10^{25}$ W~Hz$^{-1}$. Although we note that our spectral-line observations are not ideal for constructing a continuum image, these result are nevertheless in agreement with the 4.8 GHz observations by \citet[][$\ $see also Section \ref{sec:introduction}]{mor93}, when taking into account the spectral index of the radio source \citep[see][]{wal85}.

\subsection{H{\scriptsize \ }{\small I} emission}
\label{sec:emission}

\begin{figure*}
\centering
\includegraphics[width=0.8\textwidth]{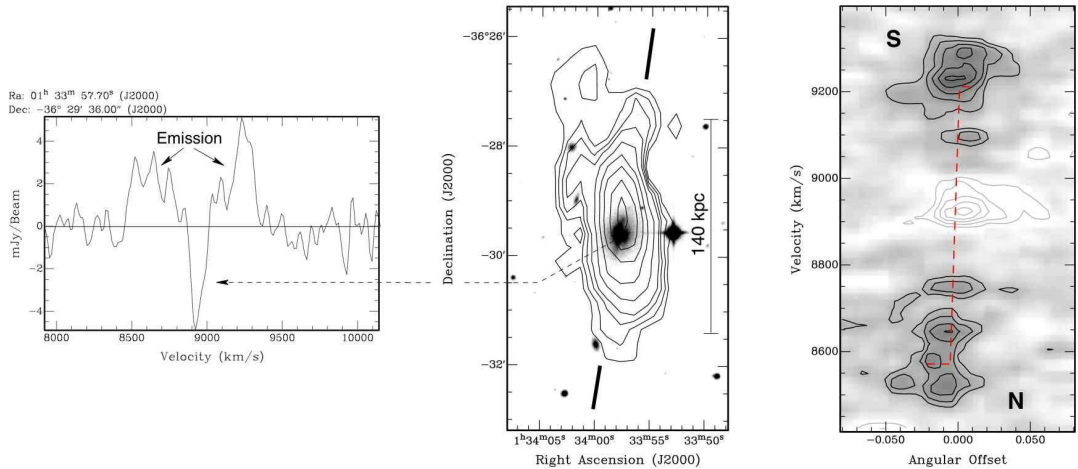}
\caption{{\sl Middle:} total intensity map of the \HI\ gas in NGC~612 (contours) overlaid on to an optical SDSS image (grey-scale). Contour levels \HI: 0.8, 1.1, 1.5, 1.9, 2.8, 4.0, 5.1, 6.5, 7.8 $\times 10^{19}$ cm$^{-2}$. {\sl Left:} \HI\ absorption profile against the central part of the radio galaxy. {\sl Right:} position-velocity plot of the \HI\ along the disc of the host galaxy (PA 171$^{\circ}$, as indicated by the broken line in the middle image). The dotted line presents the position-velocity curve derived from optical emission lines by \citet{gos80} measured along a slightly different PA (168$^{\circ}$). Contour levels \HI: -4.3, -3.6, -2.9, -2.2, -1.5 (grey); 1.5, 2.2, 2.9, 3.6, 4.3 (black) mJy beam$^{-1}$.}
\label{fig:HIemission}
\end{figure*}

\begin{figure*}
\centering
\includegraphics[width=0.78\textwidth]{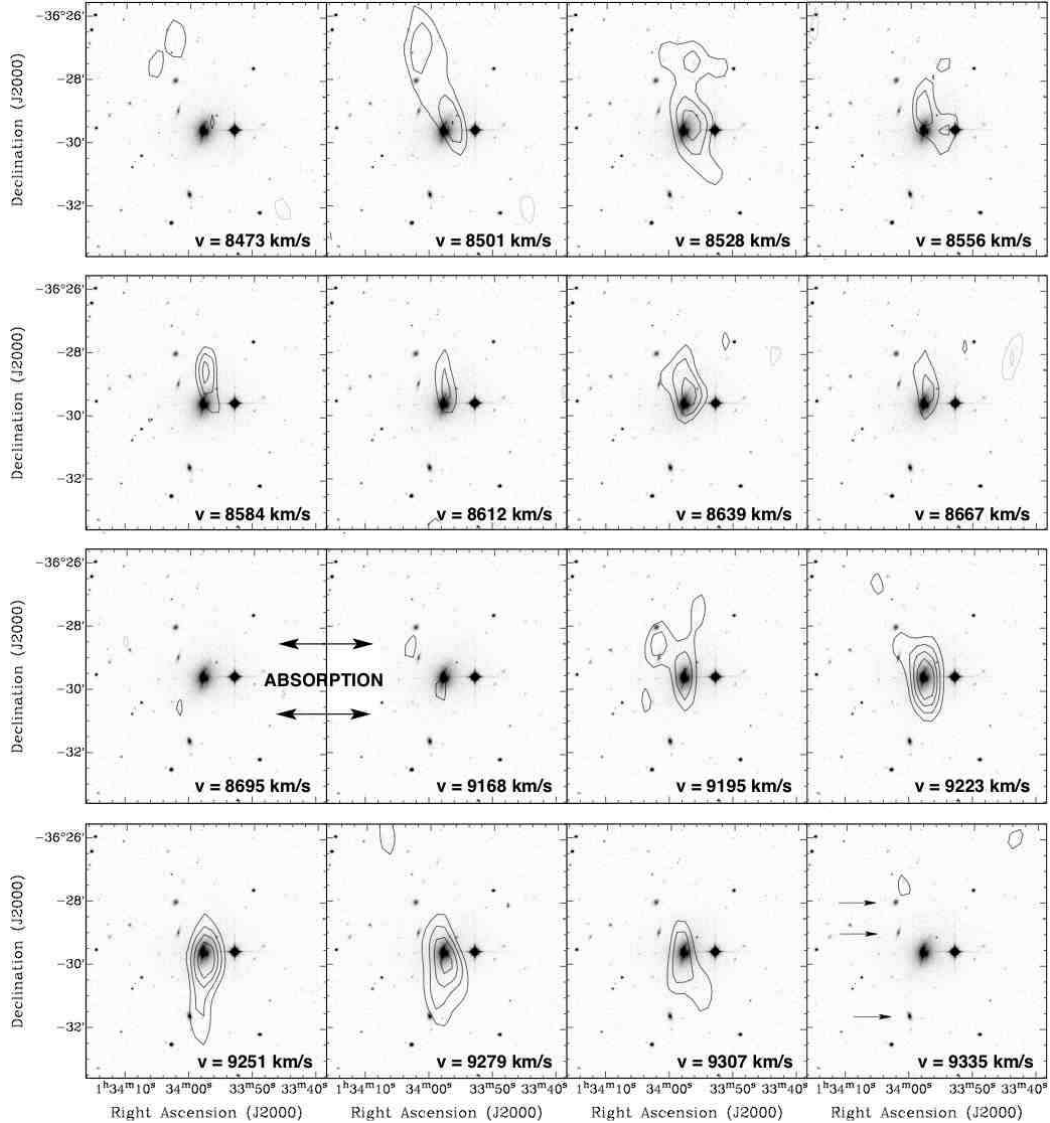}
\caption{Channel maps of the \HI\ emission in NGC~612. Contour levels are at the 3, 4, 5 and 6$\sigma$ level, corresponding to -2.5, -1.9 (grey); 1.9, 2.5, 3.2, 3.8 (black) mJy beam$^{-1}$. The three possible small companions are indicated with an arrow in the last frame (see also Fig. \ref{fig:vlt}).}
\label{fig:channelmaps}
\end{figure*}

Figure \ref{fig:HIemission} {\sl (middle)} shows a total intensity map of the \HI\ gas detected in emission in NGC~612. The total mass of this \HI\ gas is $1.8 \times 10^{9} M_{\odot}$.\footnote{All the \HI\ mass estimates given in this paper have been corrected for the primary beam of the telescope.} In the central region, \HI\ is detected in absorption against the radio continuum (Fig. \ref{fig:HIemission} - {\sl left}), hence the total \HI\ mass in NGC~612 is somewhat larger. Most of the \HI\ gas in NGC~612 is located approximately along the direction of the optical disc and dust-lane and has an average surface density at the location of the host galaxy of about 3.1 $M_{\odot}\ {\rm pc}^{-2}$. As illustrated in Figs. \ref{fig:HIemission} {\sl (right)} and \ref{fig:channelmaps}, the \HI\ gas along the optical disc of NGC~612 appears to form a rotating structure (seen edge-on) with a total extent of about 140 kpc. This \HI\ structure covers a velocity range of 850 \kms, centred on $v = 8900$ \kms. This is in good agreement with the central velocity of the \HI\ absorption, which occurs at $v = 8925$ \kms\ (see also Fig. \ref{fig:highresabs}) and which, we assume, indicates the systemic velocity of the system. The resolution of our observations is too low to determine the detailed morphology and kinematics of the \HI\ gas and hence also whether or not the bulk of the \HI\ is fully settled in regular rotation. Its kinematics nevertheless appear similar to that of the optical emission-line gas in this system, which has been observed by \citet{gos80} to follow regular rotation along the direction of the dust-lane out to about 28 kpc from the centre, covering a velocity range of 680 \kms (centred on $v = 8900 \pm 50$ \kms). 

Despite the apparent regular rotation, there is clearly some degree of asymmetry visible in the disc (the clearest asymmetries are visible in frames 2/3 and 11 of Fig. \ref{fig:channelmaps}; i.e. at v = 8501/8528 and 9195 \kms). Because of the limited spatial resolution of our observations we cannot study these asymmetries in detail. Possibly, a warp in the disc of NGC~612 could be responsible for bending the disc at $v = 8501/8528$ \kms. Alternatively, the asymmetries may be tail-like features that stretch toward three small galaxies, whose redshifts are unknown. These galaxies are located 37 and 66 kpc north-east and 76 kpc south of NGC~612 spatially and are indicated with an arrow in the last plot of Fig. \ref{fig:channelmaps}. Figure \ref{fig:vlt} shows these nearby galaxies in more detail. If these galaxies are true companions of NGC~612, the tail-like \HI\ features could indicate that interactions are ongoing between them and the disc of NGC~612.

\subsection{H{\scriptsize \ }{\small I} environment}
\label{sec:environment}

\begin{figure*}
\centering
\includegraphics[width=\textwidth]{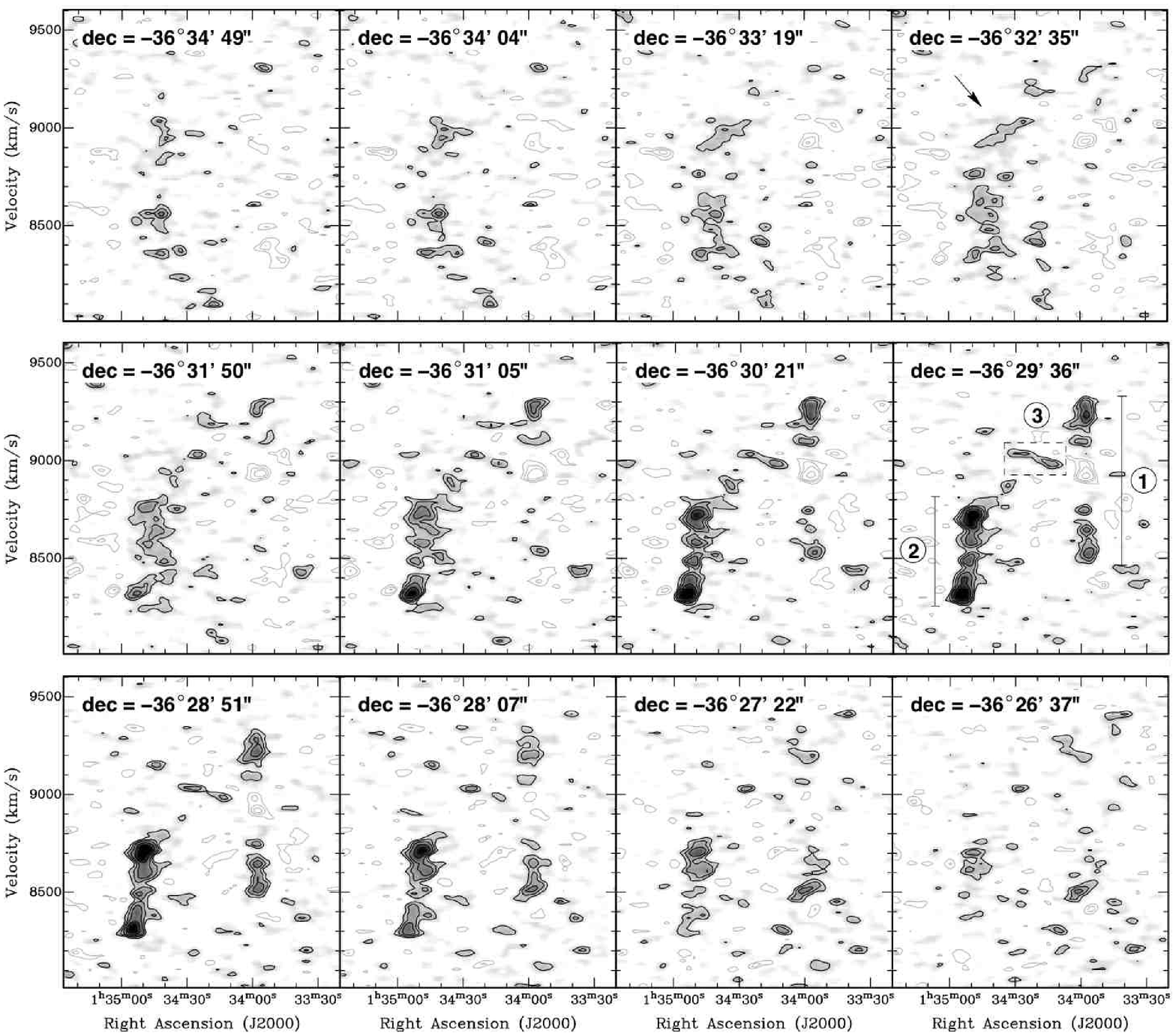}
\caption{Consecutive maps (plotted in velocity versus R.A.) of the data cube. Every channel contains the integrated signal across 44.7 arcsec (or 26.8 kpc) in Dec. Contour levels are at the 2, 3, 4 and 6$\sigma$ level (black for emission, grey for absorption). The features 1, 2 and 3, as well as the arrow, are discussed in the text and are also presented in the total intensity image of Fig. \ref{fig:envlowres}.}
\label{fig:channelmapsXZ}
\end{figure*}

\begin{figure*}
\centering
\includegraphics[width=0.79\textwidth]{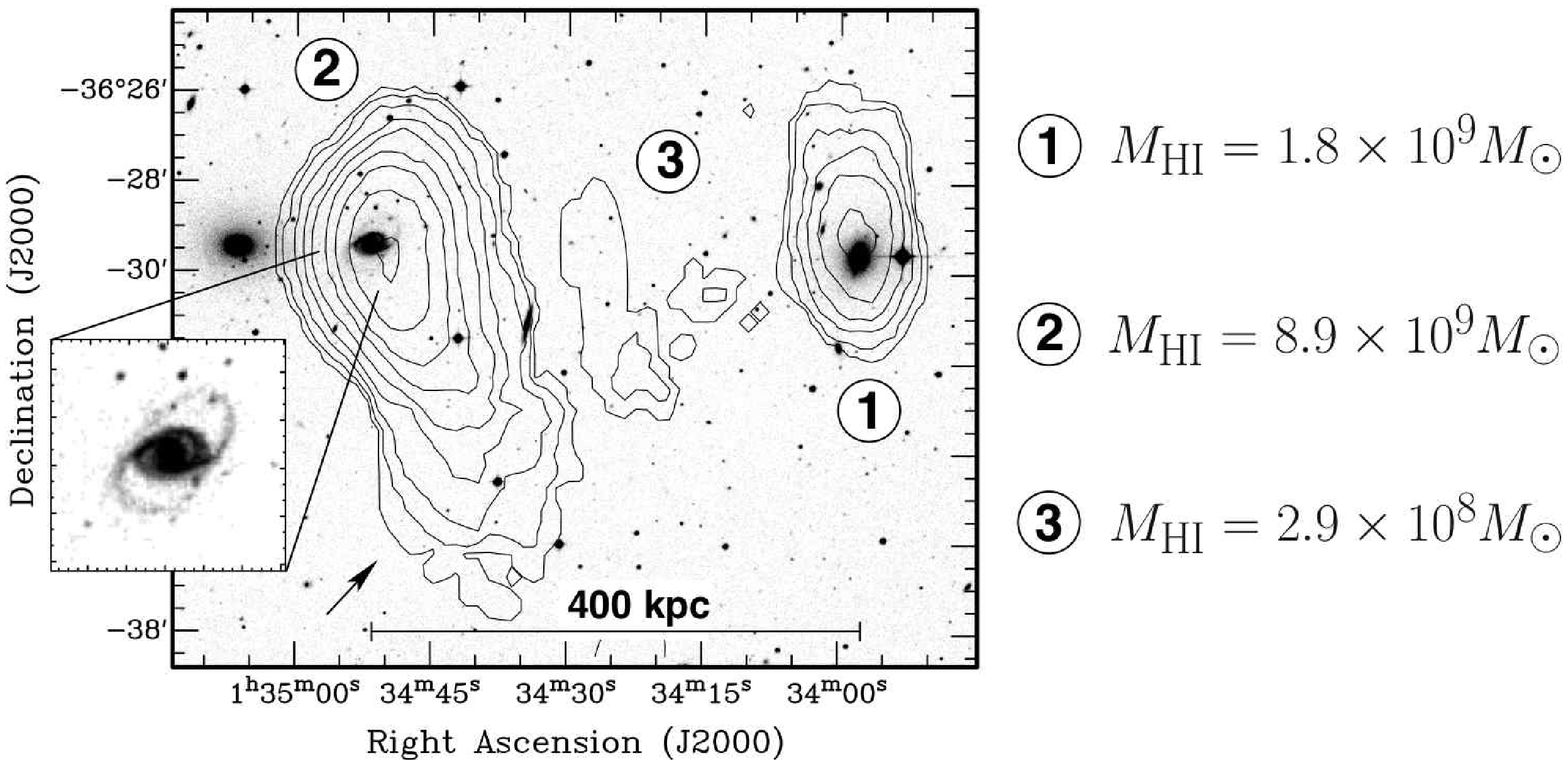}
\caption{Total intensity map of \HI\ emission from the low-resolution smoothed data cube (contours) overlaid on to an optical SDSS image (grey scale). Contour levels: 0.2, 0.4, 0.7, 1.1, 1.7, 2.3, 3.4, 4.5, 5.7 $\times 10^{19}$ cm$^{-2}$. The arrow indicates the prominent \HI\ tail, which is the same feature marked by the arrow in Fig. \ref{fig:channelmapsXZ}.}
\label{fig:envlowres}
\end{figure*}

Figure \ref{fig:envlowres} shows the total intensity \HI\ map of the larger environment of NGC~612 at lower resolution (see below). The galaxy NGC~619 is located at a projected distance of 400 kpc east of NGC~612. The `8-shaped' optical appearance of NGC~619 is typical for galaxies with a prominent bar \citep[e.g.][]{ath99}. NGC~619 contains $8.9 \times 10^{9} M_{\odot}$ of \HI. Most of this \HI\ gas is located at the position of the host galaxy. A prominent tail of \HI\ gas stretches 219 kpc toward the south-west. This \HI\ tail has no identifiable counterpart in the optical Sloan Digital Sky Survey (SDSS) image shown in Fig. \ref{fig:envlowres}.

When inspecting the data cube in the {\it xz} [right ascension (R.A.) - velocity] plane, we find clear evidence for additional, but very faint, \HI\ emission-line gas in the region between NGC~612 and NGC~619. This gets clear from Fig. \ref{fig:channelmapsXZ}. Here we show consecutive maps, plotted in velocity against R.A., with  the integrated signal across 44.7 arcsec  (or 26.8 kpc) in declination (to clarify the faint emission, the full-resolution data were smoothed in the image-domain by 10$\%$ in the spatial directions). Given the large beam-size of the observations, the consecutive plots in Fig. \ref{fig:channelmapsXZ} are not entirely independent. Nevertheless, there are a few interesting features that emerge in Fig. \ref{fig:channelmapsXZ}. The emission in regions 1 and 2 is directly related to NGC~612 and NGC~619. The \HI\ tail south-west of NGC~619 is clearly visible in the third and fourth frame (arrow). This tail connects to a faint stream of \HI\ in the direction of NGC~612 (feature `3', as indicated in the eighth frame). The faint \HI\ stream -- most likely some sort of bridge between both systems -- appears to end at the central region as well as the systemic velocity of NGC~612. Because this \HI\ bridge is detected only at about the 3$\sigma$ level and the feature is extended in {\it x} and {\it y}, but not much in {\it z}, this feature is not immediately apparent in the {\it xy} planes of the data-cube or in a total intensity image of the full resolution data. In order to give an indication of the spatial distribution of the \HI\ bridge, we smoothed our data spatially to a resolution of $218.1 \times 178.4$ arcsec$^{2}$ and we applied an additional Hanning smooth in velocity. Figure \ref{fig:envlowres} shows a total intensity map made from this smoothed data by summing all the signal above 3$\sigma$ only in the velocity range of the emission features. The \HI\ tail south-west of NGC~619 and the connecting faint stream or bridge of \HI\ in the direction of NGC~612 (feature `3') are clearly visible in this image. We note, however, that additional observations are required to accurately map the \HI\ bridge spatially.

\subsection{H{\scriptsize \ }{\small I} absorption}
\label{sec:absorption}

\begin{figure}
\centering
\includegraphics[width=0.46\textwidth]{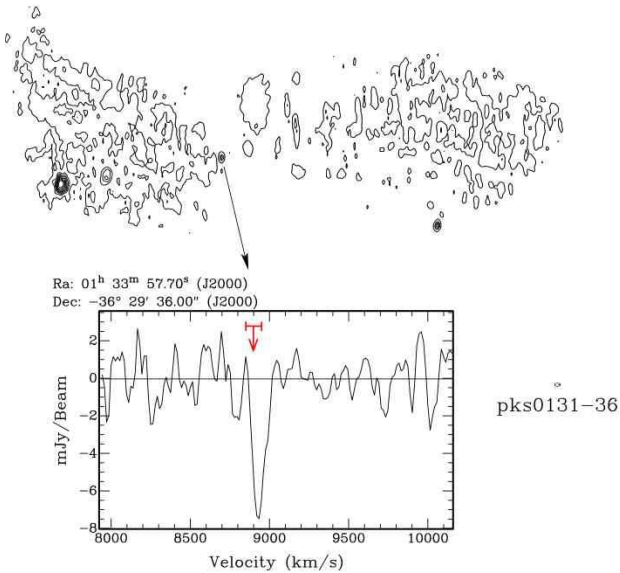}
\caption{{\sl Top:} High-resolution continuum image of PKS~0131-36 (NGC~612) including the sixth ATCA antenna. Contour levels: from $3.4 - 63$ in steps of 8.5 mJy beam$^{-1}$. {\sl Bottom:} \HI\ absorption profile detected against the unresolved radio core. For comparison, the red arrow indicates the systemic velocity derived from optical emission lines by \citet{gos80}.}
\label{fig:highresabs}
\end{figure}

\HI\ is detected in absorption at the location of the host galaxy. Figure \ref{fig:HIemission} {\sl (left)} shows the absorption profile from the data cube that we used for the \HI\ emission study (with robust weighting +1). In this data-cube, the absorption profile is diluted by \HI\ emission. In order to study the absorption in more detail, we made a high resolution data cube using the 750C data plus the sixth ATCA antenna (see Section \ref{sec:observations}). The result is shown in Fig. \ref{fig:highresabs}. The high resolution radio continuum map including the sixth antenna misses data from baselines between 750m and about 4km, which results in a large gap in uv-coverage. Because there is no information about the structure of the radio source at the spatial scales that correspond to the gap in uv-coverage, much of the extended continuum emission is resolved out in the resulting high-resolution continuum image. However, the long baselines of the sixth antenna allow us to trace the nucleus as an unresolved point-source, which is not detected in the lower resolution continuum image in Fig. \ref{fig:continuum}. By constructing also a uniform weighted data cube (in which the large-scale \HI\ emission is resolved out), we are able to recover the total flux of the \HI\ absorption feature detected in the robust weighted data (as measured from the peak of the emission in Fig. \ref{fig:HIemission}) against the very core of the radio source (Fig. \ref{fig:highresabs} - bottom). The radio continuum outside the core in Fig. \ref{fig:highresabs} is too weak to detect absorption of \HI\ gas with similar column densities. 

The absorption has an optical depth of $\tau \sim 28\%$ and full width at half-maximum (FWHM) $\sim 100$ \kms, which corresponds to a column density of $N_{\rm H{\small I}} \sim 5.1 \times 10^{21}$ cm$^{-2}$ (for $T_{\rm spin} = 100$ K). This column density is significantly larger than that derived from the lower resolution emission-line data (Fig. \ref{fig:HIemission}). However, the \HI\ absorption traces gas in our line-of-sight roughly through the mid-plane of the edge-on disc, whose scale-height is much smaller than the large beam-size of the emission-line study. Beam-smearing effects could therefore severely underestimate the true column densities of the \HI\ disc in the emission-line analysis (locally, this effect could even be enhanced if the \HI\ gas in the disc is clumpy). Alternatively, a substantial part of the absorbing gas may be located very close to the nucleus (e.g. in a nuclear disc or torus).\\
\vspace{-2mm}\\
Against the eastern radio hot-spot, no \HI\ has been detected in absorption down to a 5$\sigma$ level of $\tau \sim 0.6\%$ (for the robust +1 data). This corresponds to an upper limit in \HI\ column density of $N_{\rm HI} \la 1 \times 10^{20}$ cm$^{-2}$ (for FWHM = 100 \kms\ and $T_{\rm spin} = 100{\rm K}$) that lies in our line of sight to the radio hot-spot.

\subsection{Optical shells}
\label{sec:optical}

\begin{figure*}
\centering
\includegraphics[width=0.75\textwidth]{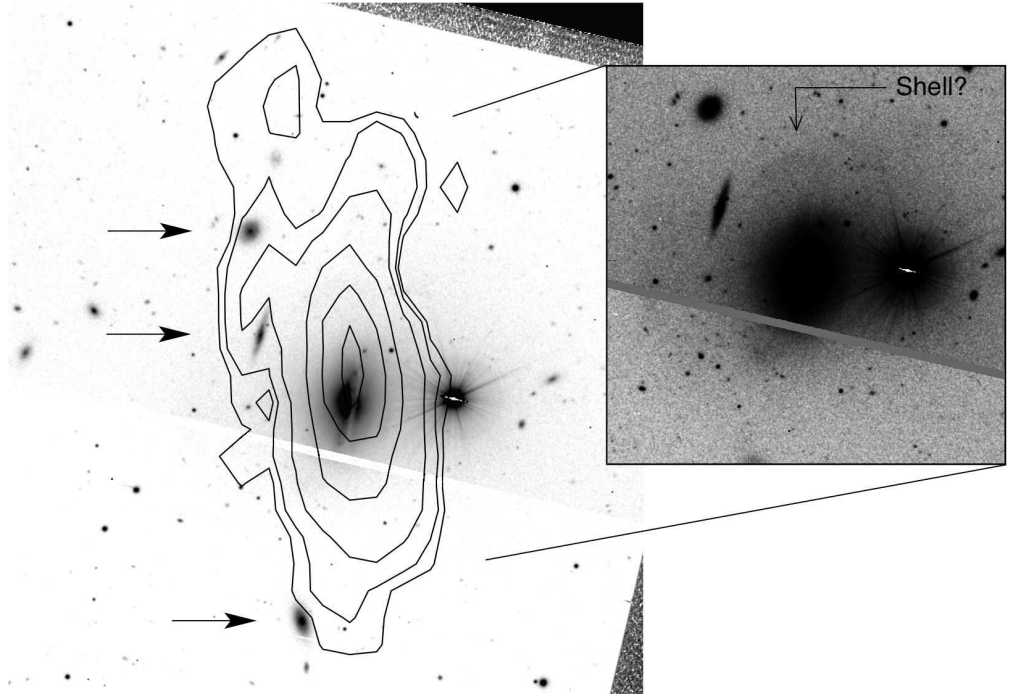}
\caption{{\sl Left:} Contours of \HI\ emission overlaid on to the optical VLT acquisition image (Section \ref{sec:VLTimage}) of NGC~612 and its direct environment. Contour levels: 0.8, 1.1, 1.9, 4.0, 6.1, 7.8 $\times 10^{19}$ cm$^{-2}$. {\sl Right:} high-contrast plot of the VLT acquisition image of NGC~612. The diagonal line across the image is the overlap region of the two chips (see Section \ref{sec:VLTimage})}
\label{fig:vlt}
\end{figure*}

Figure \ref{fig:vlt} shows the VLT acquisition image (Section \ref{sec:VLTimage}) of the environment of NGC~612. Although a high quality image of NGC~612 itself with a limited field-of-view has already been analysed by \citet{ver01}, this acquisition image is the best available image for studying the optical structure of NGC~612 at the location of the outer disc and for identifying small galaxies surrounding NGC~612. 

As mentioned in Section \ref{sec:emission}, part of the \HI\ emission in the disc of NGC~612 seems to extend in the direction of three nearby small galaxies. These three galaxies are clearly visible in Fig. \ref{fig:vlt} {\sl - left} (indicated with an arrow, as in Fig. \ref{fig:channelmaps}). A redshift determination of these galaxies is necessary to verify whether these systems are true companions of NGC~612 or background objects. In addition, many other galaxies are visible in the VLT acquisition image. 

Figure \ref{fig:vlt} {\sl (right)} shows NGC~612 in high contrast. A very faint shell-like structure is visible in this plot. The apparent optical shell is visible across the northern part of NGC~612, just west of one of the three small nearby galaxies toward which part of the \HI\ structure appears to extend. A possible very faint optical extension may also be present in the southern part of NGC~612, although this needs to be verified. Since the integration time of our acquisition image is only 10 seconds, deep optical imaging should reveal these faint optical structures in much more detail.

\section{Discussion}
\label{sec:discussion}

We detect large-scale \HI\ in the nearby radio galaxy NGC~612. The fact that NGC~612 harbours a powerful radio source with a clear hot-spot in one of the jet/lobe structures makes it one of the few powerful \FRII\ radio galaxies -- given their low volume density in the nearby Universe -- for which accurate mapping of the \HI\ emission-line gas can be done with the sensitivity of current-day radio telescopes. However, in the remainder of the discussion we would like the reader to keep in mind that, rather than being the nearby counterpart of powerful high-$z$ \FRII\ radio galaxies and quasars, the radio source properties of NGC~612 resemble more closely those found in the transition region between \FRI\ and \FRII\ sources (see Section \ref{sec:introduction}).

\subsection{Star-forming \HI\ disc}
\label{sec:HIdisc}

NGC~612 contains a large-scale, edge-on \HI\ disc, which appears to follow the optical disc and dust lane, but is observed out to a larger radius than the optical emission-line disc. The bulk of the \HI\ gas in the disc appears to be settled in regular rotation, although asymmetries in the \HI\ distribution -- likely the result of either a warp or tidal interactions with small companion galaxies -- indicate that perturbations are being exerted on part of the gas in the disc. \citet{gos80} estimate that the optical $B$-band luminosity of NGC~612 is $L_{B} = 2.0 \times 10^{11} L_{\odot}$, which means that $M_{\rm HI}/L_{B} = 0.009$ for this system (please note that this estimate does not take into account the \HI\ gas detected in absorption against the central radio continuum, hence the true value of $M_{\rm HI}/L_{B}$ could be somewhat larger). From a literature study on a heterogeneous sample, \citet{war86} derive that about 24$\%$ of nearby S0 galaxies have been detected in HI. The $M_{\rm HI}/L_{B}$ that we derive for NGC~612 is within the broad range, albeit at the low end, of the distribution for these \HI\ detected S0 galaxies from the literature. The broad distribution of $M_{\rm HI}/L_{B}$ values in S0 galaxies has led \citet{war86} to conclude that much of the cold gas in S0s, like in ellipticals, has an origin completely external to the galaxy.

The average surface density of the \HI\ disc in NGC~612 at the location of the optical host galaxy ($3.1\ M_{\odot}\ {\rm pc^{-2}}$)  is close to the critical gas surface density for star formation in galaxy discs \citep{hul93,mar01}. As a result, star formation is likely to happen across the disc, or at least at discrete locations, where the \HI\ surface density is locally higher than the average value derived from our \HI\ observations. This is in agreement with the presence of a prominent young stellar population that has been traced throughout the stellar disk by \citet{hol07} out to a radius of at least 15 kpc.

\subsubsection{Dark matter halo}
\label{sec:DM}

Assuming the \HI\ gas in the disk of NGC~612 follows regular rotation and the underlying dark-matter halo has a spherical distribution, we can make an estimate of the total mass enclosed by the system:
\begin{equation}
M_{\rm enc} = {{R_{\rm out}\ v_{\rm out}^{2}}\over{{\rm sin}^{2}i\ G}}, 
\end{equation}
with $R_{\rm out}$ the outer radius of the rotating disc, $v_{\rm out}$ the observed velocity of the disc at this distance, $i$ the inclination of the disk and $G = 6.673 \times 10^{-11}$ m$^{3}$ kg$^{-1}$ s$^{-2}$ the gravitational constant. Assuming $R_{\rm out} = 70$ kpc and $v_{\rm out} = 425$ \kms\ for NGC~612, $M_{\rm enc} = 2.9 \times 10^{12}\ {\rm sin}^{-2}i\ M_{\odot}$ (with $i$ close to 90$^{\circ}$). This is more than twice the value estimated by \citet{gos80} from the optical emission-line disc (which is much less extended than the \HI\ disc and shows a slightly lower rotational velocity). Our estimate is most likely an upper limit, because a more flattened distribution of the underlying dark-matter halo would decrease $M_{\rm enc}$. Moreover, as we mentioned before, due to the limited resolution of our observations it is not exactly clear how settled the gas is in regular rotation, in particular in the outer parts of the disc. Nevertheless, given that NGC~612 has a luminosity of $L_{B} = 2.0 \times 10^{11} L_{\odot}$ \citep{gos80}, we estimate that $M_{\rm enc}/L_{B} \leq 14.5$, which indicates that NGC~612 is likely embedded in a massive dark-matter halo. Whether or not early-type galaxies {\sl in general} contain significant amounts of dark matter is still under debate \citep[see e.g.][]{rom03,hum06}, but our derived value of $M_{\rm enc}/L_{B}$ for NGC~612 is in good agreement with $M_{\rm enc}/L_{B}$ for various early-type galaxies that also contain a large-scale rotating \HI\ structure \citep[see e.g.][$\ $and other references therein]{ber93,fra94,mor97,wei07}.

\subsubsection{Infra-red properties}
\label{sec:IR}

\begin{figure*}
\centering
\includegraphics[width=0.7\textwidth]{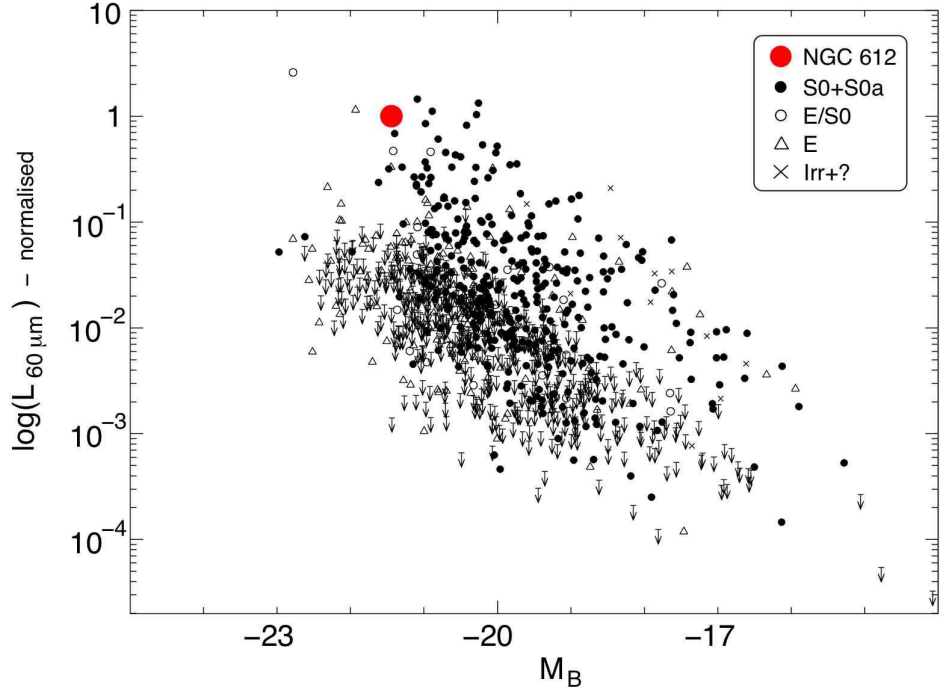}
\caption{60$\mu$m IRAS luminosity of NGC~612 compared with that of a sample of nearby early-type galaxies \citep{kna89}. The early-type galaxies are divided into S0 (and S0a) galaxies, transition objects (E/S0), ellipticals (E), and irregular/morphologically undefined galaxies (Irr+?), which are represented by the different symbols (as indicated in the plot). $L_{60{\mu}m}$ (normalised to the value of NGC~612) and $M_{B}$ are derived from $S_{60{\mu}m}$, $v$ and $m_{B}$ as given by Knapp et al. (assuming the redshift-velocity $v$ gives a good indication of the distance to the galaxies). In case of a non-detection at 60$\mu$m, a 3$\sigma$ upper limit is plotted (arrows). Galaxies for which $m_{B}$ or $v$ was not given by Knapp et al., or which have a negative redshift velocity, were not taken into account.}
\label{fig:IR}
\end{figure*}

NGC~612 also has a relatively high infra-red (IR) luminosity; from the IRAS flux measurements at 12 - 100 $\mu$m \citep{kna89} and following \citet{san96} we estimate that the total infra-red luminosity of NGC~612 $L_{\rm IR}{\rm (8-1000 \mu m)} = 4.3 \times 10^{11} L_{\odot}$, which is in the regime of Luminous Infra-Red Galaxies (LIRGs). In Fig. \ref{fig:IR} we compare the relative 60$\mu$m IRAS luminosity of NGC~612 with that of nearby early-type galaxies from a large sample studied by \citet{kna89}. We derive the 60$\mu$m luminosity assuming $L_{60{\mu}m} = 4\pi\nu S_{\nu}D^{2}/L_{\odot}$, corresponding to $L_{60{\mu}m} \approx 4 \times 10^{10} L_{\odot}$ for NGC~612. Because our derived values for $L_{60{\mu}m}$ and $M_{B}$ do not take into account K-correction, reddening effects or true distance determination (which may affect in particular some of the nearest galaxies, whose redshift velocity may be dominated by gravitational motions due to surrounding galaxies), the values of $L_{60{\mu}m}$ and $M_{B}$ should be taken with caution. Nevertheless, the point that we want to make is that the 60$\mu$m luminosity of NGC~612 is relatively high compared with that of genuine early-type galaxies (E and S0) up to distances comparable to that of NGC~612. 

It is, however, known that radio-loud early-type galaxies on average contain a significantly larger 60$\mu$m luminosity than their radio-quiet counterparts \citep{gol88,kna90}. In fact, when comparing NGC~612 with powerful radio galaxies in the nearby Universe \citep{gol88}, a significant fraction of radio-loud objects have a 60$\mu$m luminosity comparable or larger than that of NGC~612 \citep[and similar to that of spirals;][]{jon84}. Although this fraction is somewhat smaller for radio galaxies with weak emission lines (like NGC~612), the emission-line region in NGC~612 is extended, which is an indication for the presence of an appreciable ISM that contributes to the IR luminosity \citep{gol88}. It therefore seems that the IR luminosity of NGC~612 is strong compared to early-type galaxies in general, but not uncommon for powerful radio galaxies. \citet{gol88} propose that galaxy-galaxy interactions can cause the strong IR emission and simultaneously trigger the activity in powerful radio galaxies.

Although a substantial part of the 60$\mu$m emission in active galaxies often comes from a dust-enshrouded AGN, the lack of a powerful optical AGN in NGC~612 (see Section \ref{sec:introduction}), combined with its relatively cool mid-IR colour \citep{kna89}, implies that most of the 60$\mu$m emission in this system is the result of dust-heating by stars. Additional evidence for this comes from the fact that NGC~612 clearly lies above the correlation between the 70$\mu$m IR luminosity and the [\OIII]$\lambda$5007 emission-line flux due to AGN heating in powerful radio galaxies presented by \citet{tad07}. It is interesting to note, however, that the infra-red colours of NGC~612 \citep[with log$({{f_{\nu}(60\ \mu{\rm m})}\over{f_{\nu}(100\ \mu{\rm m})}})$ = -0.48 and log$({{f_{\nu}(12\ mu{\rm m})}\over{f_{\nu}(25\ mu{\rm m})}})$ = 0.20;][]{kna89} suggest that its IR flux does not originate entirely from actively star forming regions associated with the young stellar population in NGC~612 (as is often the case for IR-luminous starburst galaxies), but that an appreciable fraction likely comes from cirrus irradiated by the galaxy's total stellar population \citep[see][$\ $note that the IR colours of NGC~612 are similar to e.g. those of M81]{hel86}. Detailed IR spectroscopy is required for a more quantitative analysis of the nature of the IR emission in NGC~612.




\subsection{NGC~612, a special radio galaxy?}
\label{sec:special}

The presence of an extended \HI\ and optical emission-line disc, a young stellar population at various locations in this disc, a massive dark matter halo and a high 60$\mu$m IR luminosity suggest that NGC~612 -- despite the fact that the host galaxy has a regular S0 optical morphology \citep{ver01} -- may be considered a star-forming disc galaxy (be it with a relatively large bulge), although it is not clear to how far out in the low surface brightness \HI\ disc star formation occurs. If dust extinction is important in the relatively large column of cold gas along our line-of-sight to the edge-on \HI\ disc, it is possible that a faint optical stellar counterpart in the outer parts of this disc may become visible only in deep optical imaging across a large enough field of view.

The occurrence of an extended radio source in a host galaxy with regular optical morphology and a galaxy-scale star-forming disc is a rare phenomenon. \citet{ver01} argue that almost all classical double radio sources in the nearby Universe (FR{-\small I}s and FR{-\small II}s) are associated with elliptical galaxies, that the more disky S0 hosts are often misidentified ellipticals and that NGC~612 is therefore clearly an exception to the rule. This view is shared by \citet{led01} and \citet{kee06}, who studied in detail another exceptional case, namely the spiral host galaxy of the extended \FRI\ radio source 0313-192. Another spiral host galaxy with an \HI\ disc and a radio source that reaches outside the optical body of the galaxy is B2~0722+30 \citep{cap00,emo07}, but its \FRI\ radio source is relatively weak and small compared with NGC~612 and 0313-192 and we will present a detailed analysis of this galaxy in a forthcoming paper.
 
Although the radio source in NGC~612 (with its clear \FRII\ morphology of the eastern lobe) is an order of magnitude more powerful than the edge-darkened \FRI\ radio source in the spiral galaxy 0313-192 \citep{led01,kee06}, both radio galaxies share some striking similarities: NGC~612 and 0313-192 both contain a relatively luminous bulge compared with typical spiral galaxies, which generally reflects the presence of a relatively massive central black-hole \citep[e.g.][]{kor95,geb00,fer00}. Both galaxies also show indications that tidal encounters may have occurred, which could have resulted in enhanced fuelling of the central engine (see Section \ref{sec:interactions} for more details). In addition, in both NGC~612 and 0313-192 the radio jets are aligned almost perpendicular to the disc. A high-resolution continuum image of NGC~612 by \citet{mor93} shows that the radio axis lies about 25$^{\circ}$ from the minor axis of the host galaxy in the plane of the sky. The radio source has a somewhat distorted structure outside the optical host galaxy, indicating that the source changed its direction over its lifetime and that the alignment of the radio axis could have been even closer to the minor axis at an earlier stage. It is likely that this is the direction in which the radio jets encounter the least resistance from the ISM. A massive black-hole, enhanced fuelling and/or minor resistance from the ISM might explain the presence of an extended radio source in both NGC~612 and 0313-192.

\subsubsection{Hybrid radio source morphology}
\label{sec:hybrid}

\citet{gop00} suggest that the radio source in NGC~612 has a rare hybrid morphology (see Section \ref{sec:introduction}), in which case the western jet/lobe should be classified as \FRI-like. The mere existence of hybrid radio sources has led \citet{gop00} to argue that the FR-dichotomy is likely related to interactions between the radio jet and the ambient medium on kpc scales, rather than to intrinsic properties of the central engine. The presence of a large-scale disc of neutral hydrogen in the host galaxy of NGC~612 make this an interesting object for studying this possibility in more detail, but this is beyond the scope of this paper.

\subsection{Galaxy interactions and radio source triggering}
\label{sec:interactions}

As mentioned in Section \ref{sec:introduction}, a significant fraction of the host galaxies of powerful \FRII\ radio sources show signs of a past galaxy merger or interaction event in the form of a peculiar optical morphology or the characteristics of extended emission-line regions. In these studies, it is suggested that such an event may be related to the triggering of the powerful radio source. Also for NGC~612, our study indicates that galaxy collisions or interactions likely occurred, possibly related to the triggering of the radio source and the star formation in this early-type galaxy.

Perhaps the most convincing evidence for a past galactic encounter comes from the long \HI\ tail/bridge structure between NGC~612 and NGC~619, which suggests that an interaction or collision likely took place between these two galaxies. It is interesting to note that in the vicinity of NGC~612, the \HI\ bridge is aligned in the same direction as the radio jets.

Figures \ref{fig:channelmaps} and \ref{fig:vlt} show that more recent tidal interactions could also be ongoing between NGC~612 and several small nearby galaxies, although (as mentioned in Section \ref{sec:emission}) independent redshift determinations of the small nearby galaxies are necessary to verify whether they are true companions or background objects. If confirmed, such tidal interactions are likely to induce perturbations in the disc of NGC~612, which could result in some of the observed asymmetries in the disc of NGC~612 and perhaps also the triggering of star formation and fuelling of the AGN.

Another possible scenario is that NGC~612 and its large-scale \HI\ disk formed as a result of a major merger event (i.e. a merger between two roughly equal mass galaxies, of which at least one -- and possibly both -- were gas-rich), which may also have triggered the powerful radio source. It is well established from numerical simulations that the formation of early-type galaxies can be explained by galaxy mergers with a wide range in parameters \citep[e.g.][]{too72,bek98,naa99}. A major merger may also trigger a burst of star formation, which in the dusty environment of the merging galaxies enhances the system's infra-red luminosity \citep[e.g.][]{mih94}. Simulations by \citet{dim07} show that different orbital parameters of the merging galaxies have a different effect on the tidal disruption of gaseous and stellar discs and that in particular direct encounters often create large gaseous tidal-tails. \citet{bar02} shows that such expelled tidal material can -- on time-scales of one to several Gyr -- be re-accreted on to the newly formed host galaxy and settle into a large-scale gas disc. Observationally, two recent detailed case-studies show that a large-scale, low-surface brightness \HI\ disc can indeed form around an early-type host galaxy as a result of a major merger \citep{emo06,ser06}.

For NGC~612, the early-type optical appearance of the host galaxy with a faint optical shell, the high infra-red luminosity and the presence of a massive, large-scale, low surface brightness \HI\ disc indicate that a major merger event could have occurred also in this system, although additional observations are necessary to confirm this and to determine the fate of the gas during such a possible merger event. However, {\sl if} a major merger formed NGC~612, it likely occurred several rotational periods (i.e. $>$Gyr) ago, after which the large-scale \HI\ gas had time to settle in a regular rotating disc and the host galaxy to gain its primary early-type appearance. Such a time-scale is long compared to the typical spectral ages (10$^{6} - 10^{7}$ yr) derived for extended \FRII\ radio sources \citep[e.g.][]{ale87,lea89,liu92,par02}, although \citet{par02} and \citet{blu00} argue that the ages could be as much as $0.1 - 1$ Gyr as a result of an erroneous approximation of the magnetic field in the traditional spectral ageing arguments, mixing of old and young electrons by back-flow and continuous replenishment of energetic electrons. In addition, for radio sources in major merger systems, it is not uncommon that there is a significant time-delay between the initial merger and the onset of the most recent period of radio-AGN activity \citep[][]{tad05,emo06}. Therefore, the possibility that a merger event may have triggered the radio-AGN activity in NGC~612 -- as is often suggested for other powerful \FRII\ radio galaxies -- appears to be a viable scenario, but this needs to be verified with future observations.

An intriguing possibility is that the \HI\ disc around NGC~612 may continuously accrete gas from either the large-scale \HI\ bridge between NGC~612 and NGC~619 or the small nearby companions. This could explain why the disc in NGC~612 is relatively large and massive, for example compared to the case of the nearby radio galaxy Centaurus~A (Cen~A), which -- as we will describe in detail below -- contains a much smaller disc as a likely result of an unequal mass merger \citep{gor90,sch94}.

\subsection{Large-scale \HI\ in radio galaxies}

As mentioned in Section \ref{sec:introduction}, recent studies reveal that large-scale, massive \HI\ structures -- similar to the one in NGC~612 -- are detected in a significant fraction of nearby field early-type galaxies \citep{sad02,mor06b,oos07}. But also around a number of nearby radio-loud early-type galaxies, large-scale \HI\ structures are known to exist: 

Centaurus~A, the nearest radio galaxy in the Universe, contains an \HI\ disc of $M_{\rm HI} \approx 4.5 \times 10^{8} M_{\odot}$ out to a radius of about 8 kpc from the centre \citep[][$\ $for a distance of 3.5 Mpc to Cen~A]{gor90,sch94}. It also has faint shells with $1.5 \times 10^{8} M_{\odot}$ of \HI\ gas out to about 15 kpc from the nucleus, which might represent a partial ring that rotates in the same sense as the body of the host galaxy, possibly as a result of an unequal mass merger between a large elliptical and a smaller gas-rich companion \citep{sch94}. The prominent galaxy-scale dust-disc in Cen~A, combined with two extended radio jets/lobes perpendicular to this dust-disc, has often led to the comparison with NGC~612 \citep{wes66,eke78}. For Cen~A, the total \HI\ mass is low and the extent of the \HI\ features is small compared with NGC~612 (with the bulk of the \HI\ gas in Cen~A located within the optical body of the host galaxy). Also the radio source in Cen~A, generally classified as \FRI, is almost an order of magnitude less powerful than the one in NGC~612. 

Another southern radio galaxy that contains a large-scale \HI\ disc ($M_{\rm HI} = 1.5 \times 10^{10} M_{\odot}$ and diameter = 127 kpc for $H_{0} = 71$ km~s$^{-1}$~Mpc$^{-1}$) is the early-type galaxy PKS~1718-649 \citep{ver95}. The gas in the large-scale disc of this radio galaxy is perturbed and V\'{e}ron-Cetty et al. propose that this is perhaps the result of a merger that involved at least one spiral galaxy. The radio source in PKS~1718-649 is compact and classified as a gigahertz-peaked spectrum (GPS) radio source \citep{tin03}.

\begin{figure}
\centering
\includegraphics[width=0.46\textwidth]{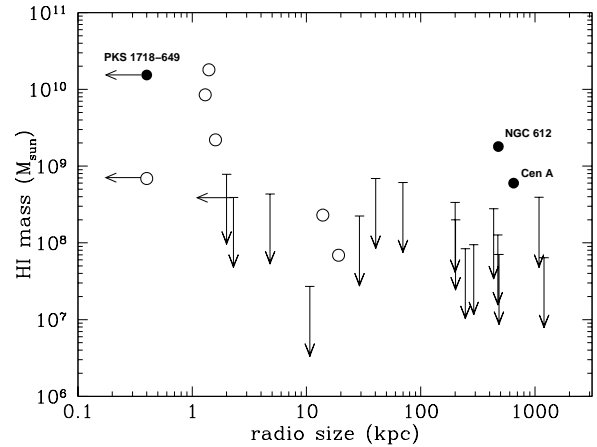}
\caption{{\sl - from \citet{emo07}}: Plot showing the total \HI\ mass detected in emission plotted against the total linear extent of the radio source for our complete sample of B2 radio galaxies (open circles and upper limits) plus the three southern radio galaxies discussed in this paper (filled circles). In case of non detection in the B2 sample, a conservative upper limit (3$\sigma$ across 400 \kms) is given. See \citet{emo07} for more details.}
\label{fig:masssizeplot}
\end{figure}
In order to investigate the large-scale \HI\ properties of radio-loud early-type galaxies in a more systematic way, we recently concluded an \HI\ study of a complete sample of nearby, non-cluster radio galaxies from the B2 catalogue \citep{emo07,emo06thesis}. The radio sources in these systems, all with radio power lower than NGC~612, have either a compact or a typical edge-darkened \FRI\ morphology and their host galaxies were a priori classified as genuine early-type systems. When comparing the \HI-properties of these B2 radio galaxies with those of NGC~612, there are some remarkable results; similar to the case of NGC~612, regular rotating large-scale \HI\ discs and rings (with $M_{\rm HI} \approx {\rm several} \times 10^{8} - 10^{10} M_{\odot}$ and diameters of several tens to hundreds of kpc) have been detected around roughly 50$\%$ of the radio galaxies that host a compact ($< 15$ kpc) radio source, while none of the early-type host galaxies of the extended \FRI\ radio sources shows any \HI\ in emission above a conservative detection limit of a ${\rm few} \times 10^{8} M_{\odot}$. This is visualized in Fig. \ref{fig:masssizeplot}, which is a plot from \citet{emo07} that shows the total \HI\ mass detected in emission around NGC~612 and the B2 radio galaxies plotted against the size of the radio source (for comparison, we also included Cen~A and PKS~1718-649). The large-scale \HI\ disk in NGC~612 morphologically resembles that of the \HI-rich compact sources \citep[although the total \HI\ mass is still about an order of magnitude lower than that of the most massive of the \HI\ discs found around the compact sources;][]{emo06,emo07}. NGC~612 contains significantly larger amounts of extended \HI\ than the \FRI\ sources in our B2 sample. 


The very nearby \FRI\ radio galaxy Cen~A (3.5 Mpc) appears to be an exceptional case regarding its \HI\ content compared with the \FRI\ sources from our B2 sample. However, we note that the bulk of the detected \HI\ emission in Cen~A is - in contrast to most of our \HI\ detections - located in a disk well inside the optical body of the host galaxy and part of this \HI\ disk would have been observed in absorption against the central radio continuum instead of emission at the typical distance of most of our B2 sources ($60-160$ Mpc); see also \citet{emo07}. A detailed analysis will be postponed to a forthcoming paper.  

We are currently conducting a study to map \HI\ gas in a small but complete sample of the nearest \FRII\ radio galaxies in order to verify whether there is a fundamental difference in large-scale \HI\ content between nearby \FRI\ and \FRII\ radio galaxies. Preliminary results appear to indicate that this may indeed be the case \citep{emo08proc}, but final results on this will be presented in a future paper.

\section{Conclusions}
\label{sec:conclusions}

We presented results of a morphological and kinematical study of the neutral hydrogen gas in and around the nearby powerful radio galaxy NGC~612 (PKS~0131-36). We compared this with information about the stellar populations, optical morphology, infra-red characteristics and radio-source properties of this radio-loud early-type galaxy. The most important conclusions that we discussed in this paper are:
\begin{itemize}
\item{The S0 host galaxy NGC~612 contains a large-scale \HI\ disc, with an \HI\ mass of at least $1.8 \times 10^{9} M_{\odot}$ and a diameter of about 140 kpc. The bulk of the \HI\ gas appears to be in regular rotation, although asymmetries in the \HI\ disc indicate that perturbations are being exerted on part of the gas, possibly by a number of nearby companions;}\\
\item{The total velocity range of the gas in the \HI\ disc is 850 \kms, which indicates that NGC~612 contains a massive dark matter halo;}\\
\item{NGC~612 is a rare example of a galaxy with an extended radio source and a large-scale star-forming disc;}\\
\item{A long, faint \HI\ bridge connects NGC~612 with the \HI-rich barred galaxy NGC~619 about 400 kpc away, possibly as the result of a past interaction between both galaxies;}\\
\item{We argue that -- in agreement with what is generally observed for powerful \FRII\ radio galaxies -- the triggering of the powerful radio source in NGC~612 could be related to ongoing or past galaxy interactions, or to a major merger event, which also may have formed the large-scale \HI\ disk in NGC~612;}\\
\item{The large-scale HI properties of \FRII\ radio galaxy NGC~612 are similar to those of several nearby compact radio sources, but different from those of nearby \FRI\ sources.}
\end{itemize}
Future studies are necessary to investigate in more detail the formation history of NGC~612 and compare this with the triggering and properties of its powerful radio source. 

\section*{Acknowledgements}

We would like to thank Jacqueline van Gorkom for useful discussions and Bob Sault and Mark Dijkstra for their help related to the observations. BE acknowledges the Netherlands Organisation for Scientific Research (NWO) for funding this project under Rubicon grant 680.50.0508. The Australia Telescope is funded by the Commonwealth of Australia for operation as a National Facility managed by CSIRO.

\bibliographystyle{mn2e} 
\bibliography{auth_total_NGC612} 

\label{lastpage}

\end{document}